# Epitaxial growth and magneto-transport properties of kagome metal FeGe thin films

Xiaoyue Song,[1,2] Yanshen Chen,[1,2] Yongcheng Deng,[1,3] 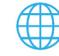 Tongao Sun,[1] Fei Wang,[4] 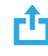 Guodong Wei,[2,a]
Xionghua Liu,[1,3,a]  and Kaiyou Wang[1,3,a]

## AFFILIATIONS

[1]State Key Laboratory of Semiconductor Physics and Chip Technologies, Institute of Semiconductors, Chinese Academy of Sciences, Beijing 100083, China
[2]School of Integrated Circuit Science and Engineering, Fert Beijing Institute, Beihang University, Beijing 100191, China
[3]College of Materials Science and Optoelectronics Engineering, University of Chinese Academy of Sciences, Beijing 100049, China
[4]Key Laboratory of Magnetic Molecules and Magnetic Information Materials of Ministry of Education & School of Materials Science and Engineering of Shanxi Normal University, Taiyuan 030006, China

[a]Authors to whom correspondence should be addressed: jellwei@buaa.edu.cn; xionghualiu@semi.ac.cn; and kywang@semi.ac.cn

## ABSTRACT

Antiferromagnetic kagome metal FeGe has attracted tremendous attention in condensed matter physics due to the charge density wave (CDW) being well below its magnetic transition temperature. Up to now, numerous works on kagome FeGe have been based on single crystal bulk, but its thin film form has still not been reported. Here, we achieved epitaxial growth of FeGe thin films on $Al_2O_3$ substrates using molecular beam epitaxy. Structural characterization with x-ray diffraction, atomic force microscopy, and high-resolution scanning transmission electron microscopy reveals single phase with flat surface of kagome FeGe thin films. Moreover, a Néel temperature of 397 K and a rapid variation of Hall coefficient and magnetoresistance around 100 K, which might be related to the CDW, were revealed via transport measurements. The high quality kagome FeGe thin films are expected to provide a versatile platform to study the mechanism of CDW and explore the application of FeGe in antiferromagnetic spintronics.

Published under an exclusive license by AIP Publishing. https://doi.org/10.1063/5.0310077

The kagome lattice, composed of hexagons and triangles in a two-dimensional (2D) network of corner-shared triangles, naturally hosts Dirac fermions, flat bands, and van Hove singularities in its electronic structure;[1–3] hence it has become a fruitful playground to study the quantum interactions between geometry, topology, spin, and correlation.[4,5] A large number of interesting phenomena have been discovered in kagome materials, including Wigner crystallization,[6] unconventional superconductivity,[7] quantum spin liquid,[8] anomalous Hall effect,[9] and topological semimetals.[10,11] Recently, the charge density wave (CDW) states in kagome superconductor $AV_3Sb_5$ (A = K, Rb, Cs) have gained tremendous attention owing to diverse CDW patterns,[12–14] sophisticated symmetry breaking,[15,16] and intertwining orders.[17] Remarkably, the antiferromagnetic kagome metal FeGe has been reported to exhibit a 2 × 2 CDW order,[18,19] where the emergence of CDW well below the magnetic transition temperature represents an exceptionally rare phenomenon, suggesting a novel correlation-driven CDW mechanism in FeGe. Although different theories have been proposed,[20–22] the origin of CDW state and how the charge order interacts with antiferromagnetism in FeGe remains controversial.

The antiferromagnetic kagome FeGe features magnetic ordering with Fe moments ferromagnetically aligned within each kagome plane but antiferromagnetically coupled along the c-axis [inset of Fig. 1(a)] with Néel temperature $T_N \sim 410$ K.[23] Below the CDW ordering temperature $T_{CDW} \sim 110$ K, concomitant anomalous Hall effects and enhanced spin polarization are found.[18] However, the out-of-plane antiferromagnetic arrangement of the Fe-kagome layers will persist down to approximately 60 K, at which point the spin texture switches from a collinear A-type antiferromagnetic to a canted double-cone antiferromagnetic structure, with an additional reorientation occurring around $\sim 30$ K.[18,24,25] Moreover, antiferromagnets have attracted extensive interest in spintronics as one of the active materials for next-generation spintronic devices, with the prospect of offering high density memory integration and ultrafast data processing.[26–31] Compared to the previously studied antiferromagnetic kagome Weyl semimetal $Mn_3Sn$ and Dirac semimetal FeSn in which the spins align in their kagome layers,[32,33] the FeGe possessing out-of-plane spin arrangement of the kagome layer with high $T_N$ would have greater application potential in antiferromagnetic spintronics.[26–28,34,35]

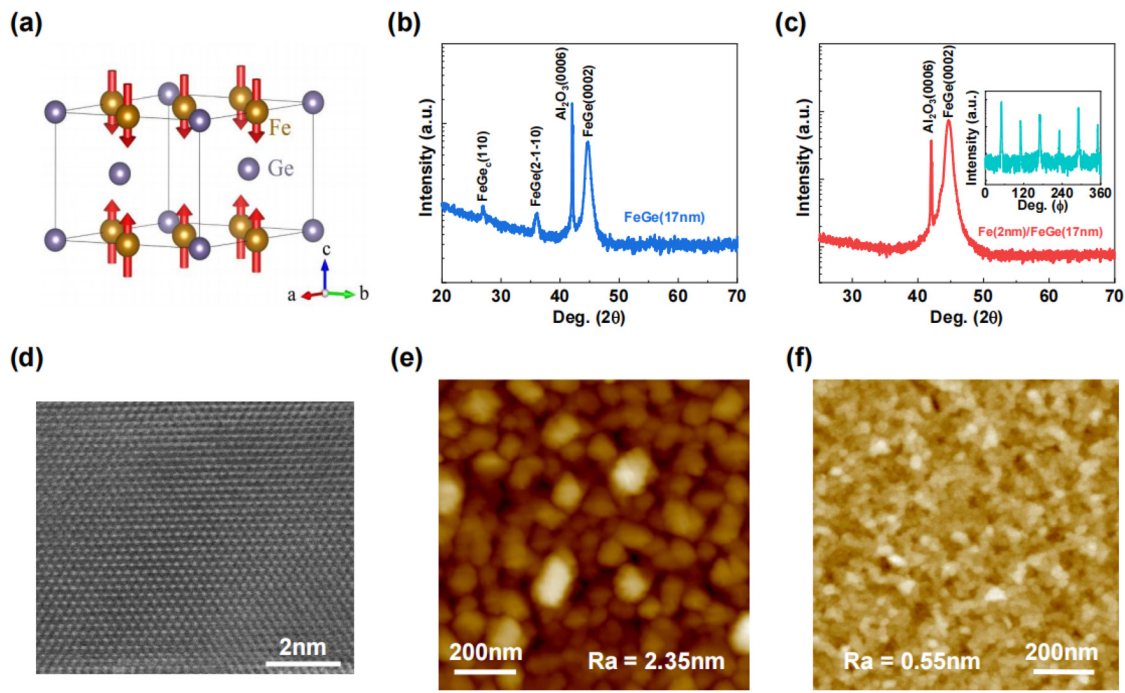

FIG. 1. (a) FeGe spin/lattice structure at room temperature. It consists of an alternating stack of 2D Fe-kagome layers and 2D Sn honey-comb layers antiferromagnetically coupled along the c-axis. (b) and (c) The x-ray diffraction patterns for 17 nm-thick FeGe thin film without (b) and with (c) 2 nm-thick Fe buffer layer. Inset of (c): $\phi$ scans conducted at a fixed angle reveal six distinct in-plane (2-1-10) Bragg diffraction peaks, with adjacent peaks separated by uniform 60° intervals. (d) The cross-sectional STEM image of FeGe thin film when viewing along the [0002] direction. (e) and (f) AFM images of a FeGe (17 nm) (e) and Fe (2 nm)/FeGe (17 nm) (f) thin films, the roughness is around 2.35 and 0.55 nm, respectively.

Although considerable work on kagome FeGe has been based on its single crystal bulk,[18–25,36] the growth of kagome FeGe thin films has still not been reported. In this work, we achieved high quality kagome FeGe thin films using molecular beam epitaxy, characterized their microstructures with x-ray diffraction, atomic force microscopy, and high-resolution scanning transmission electron microscopy, and measured the corresponding magneto-transport properties employing a physical property measurement system. The high quality kagome FeGe thin films enable us to manipulate its physical properties via strain, electrical, magnetic, or optical approaches more easily, which would help us to further understand the relation between CDW and magnetism and to explore potential applications of FeGe in antiferromagnetic spintronics.

Kagome FeGe thin films were grown by molecular beam epitaxy (MBE) in an ultrahigh vacuum (UHV) chamber with a base pressure of $5 \times 10^{-11}$ Torr. High purity Fe (99.99%) and Ge (99.9999%) were evaporated from LUXEL Radak effusion cells at temperatures of about 1290 and 1200 °C, respectively, onto single crystalline $Al_2O_3$ (0001) substrates. The Fe and Sn fluxes were calibrated using a quartz-crystal monitor (QCM) at the growth position and set to 0.70 and 0.68 Å min$^{-1}$, respectively. The substrates were annealed for 1 h at 600 °C in an oxygen pressure of $1.2 \times 10^{-2}$ Torr to gain a clean and well-ordered surface structure prior to FeGe deposition. We used a three-step growth process to obtain FeGe thin films.[37] First, a 2 nm FeGe or sub/Fe (2 nm)/FeGe (2 nm) layer was grown at 460 °C as a seed layer. Second, the film was rapidly cooled down to 100 °C and an additional 15 nm FeGe was grown at this temperature. At last, the sample was warmed up to 390 °C and annealed for 2 h to improve the crystallinity. The surface topography and phase of the thin films were checked by atomic force microscopy (AFM) and x-ray diffraction (XRD). The sample for cross-sectional scanning transmission electron microscopy (STEM) characterization was prepared using a Zeiss Auriga focused ion beam system. The samples were patterned into six-terminal Hall bar devices with a current channel width of 10 $\mu$m using standard photolithography and Ar-ion etching.[38] The magnetic properties our samples were measured using a superconducting quantum interference device (SQUID), while the magneto-transport properties were measured using a physical property measurement system (PPMS) and a low-field magneto-electrical measurement system.

We first performed microstructural characterization of our FeGe thin films. The XRD patterns of FeGe thin films are presented in Figs. 1(b) and 1(c). For the thin film directly grown on $Al_2O_3$ (0001) [Fig. 1(b)], besides the substrate $Al_2O_3$ (0006) and hexagonal FeGe (0002), other peaks, FeGe (2-1-10) and FeGe$_c$ (110), were also observed, meaning the minor impurity phase such as cubic FeGe with ferromagnetism was included.[39] For the thin film grown on 2 nm Fe buffer layer [Fig. 1(c)], the impurity peaks disappear except for FeGe (0002). Notably, Fe atoms in the (110) plane assemble into a distorted hexagonal configuration with four sides of length $s_1 = 2.48$ Å and two sides of length $s_2 = 2.87$ Å;[40] this structural arrangement offers an idea buffer layer for the epitaxial growth of FeGe thin films. Since the diffraction peak of Fe (110) is positioned at around 45°, very close to that of FeGe (0002), we were unable to resolve these two peaks individually. Furthermore, we performed in-plane XRD measurements, as presented in the inset of Fig. 1(c). The $\phi$ scans along a fixed angle reveal six distinct in-plane (2-1-10) Bragg peaks, with adjacent peaks separated by evenly spaced 60° intervals, which originate from the FeGe layer in our Fe(2 nm)/FeGe(17 nm) sample. This indicates the introduction of a Fe buffer layer is highly beneficial to the formation of hexagonal FeGe thin film.

AFM is used to examine the surface morphology of our thin films. As presented in Fig. 1(f), a continuous film with roughness of ∼0.55 nm in Fe (2 nm)/FeGe (17 nm) was observed, whereas FeGe (17 nm) without Fe buffer layer displays granular film with roughness of about 2.35 nm [inset of Fig. 1(e)]. The results indicate the Fe buffer layer can also greatly enhance the flatness of the FeGe thin film. Figure 1(d) exhibits the atomic-resolution STEM image of the thin film when viewed along the [0002] direction, revealing the expected hexagonal periodicity that matches the projected atomic arrangement.[41] These characterizations confirm the high quality of our FeGe thin films.

We then measured the transport properties of our Fe (2 nm)/FeGe (17 nm) thin film. The longitudinal resistivity vs temperature, $\rho_{xx}$ vs $T$, curve under zero field exhibits an extremum value at 397 K, as the arrow indicates in Fig. 2(a), which should represent $T_N$ of FeGe thin films, slightly lower than that of the single crystal bulk, 410 K.[21] This difference is rational and probably arises from finite-size effects in the epitaxial film. Upon cooling the sample, $\rho_{xx}$ gradually reduces with temperature, similar to the typical electrical behavior of FeGe bulk,[18,25,36] and we did not observe a clear drop in resistivity around 100 K, suggesting the first-order structural phase transition is very weak in our thin film. However, the $d\rho_{xx}/T$ vs $T$ curve presents a clear kink variation from 100 K [see the dashed line in the upper-left inset of Fig. 2(a)], which may be indicative of the CDW transition. As the resistance of 2 nm-thick Fe buffer layer is much larger than that of the whole Fe (2 nm)/FeGe (17 nm) thin film, here we considered that the $\rho_{xx}$ vs $T$ curve under zero magnetic field could mostly reflect the transport behavior of FeGe thin film. Moreover, we measured the temperature-dependent magnetization ($M$–$T$ curve spanning 5–400 K) under an applied field of 0.02 T following zero-field cooling of the sample from 300 K, see the lower-right inset of Fig. 2(a). The M–T curve exhibits no discernible variation at approximately 397 K, a phenomenon arising from the fact that the antiferromagnetic signal of FeGe is overwhelmed by the stronger magnetic contribution from the Fe layer.

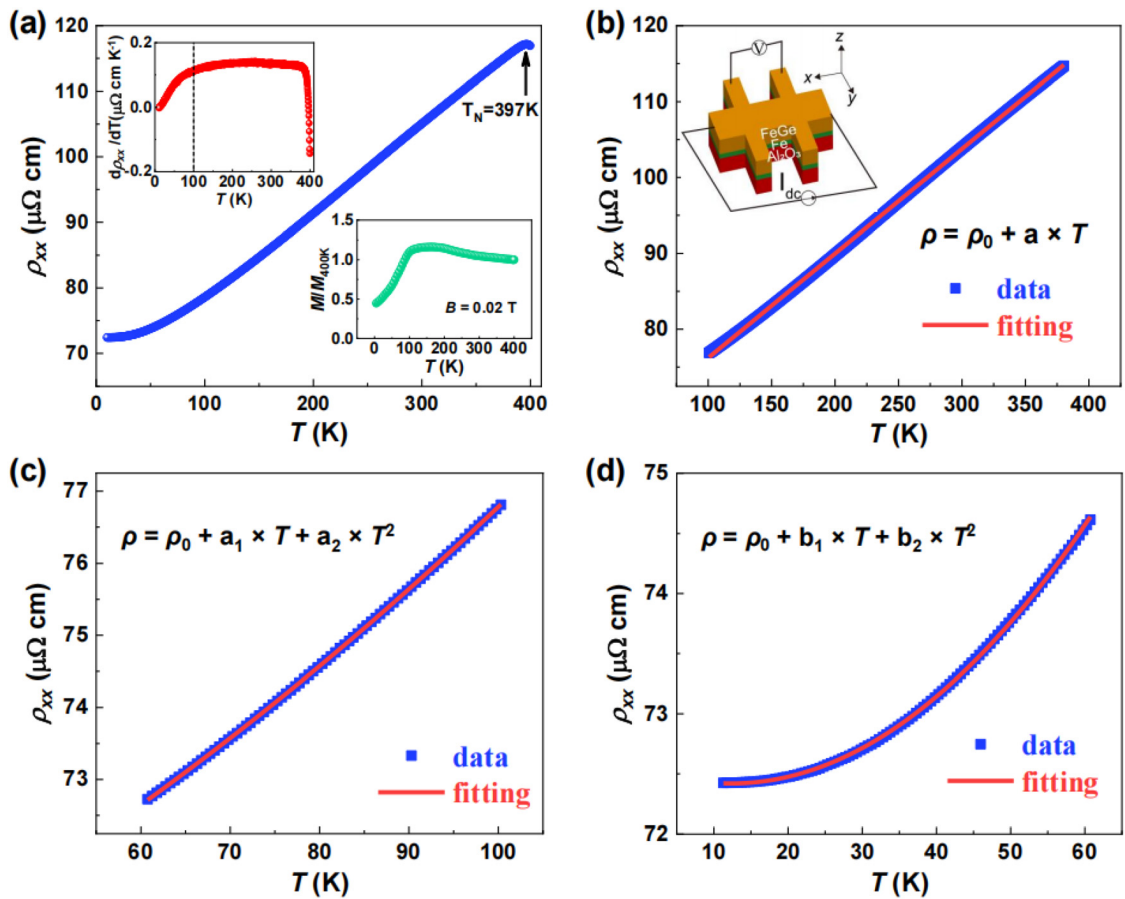

FIG. 2. (a) Temperature dependence of resistivity $\rho_{xx}$ under zero magnetic field from 2 to 400 K for Fe (2 nm)/FeGe (17 nm) thin film. Inset: (upper left) the differentiation of resistivity $d\rho_{xx}/dT$, and (lower right) the normalized temperature-dependent magnetization measured under an applied field of 0.02 T after zero-field cooling from 300 K. The Néel temperature of the FeGe thin film is about 397 K as the arrow indicates. (b)–(d) The $\rho_{xx}$ vs $T$ curve was fitted by different equations in three temperature ranges: $\rho = \rho_0 + a \times T$ for 100–380 K (b) (inset: a schematic diagram of the device testing setup), $\rho = \rho_0 + a_1 \times T + a_2 \times T^2$ for 60–100 K (c), and $\rho = \rho_0 + b_1 \times T + b_2 \times T^2$ for 10–60 K (d). The EPS plays a dominant role at high temperatures, 100–380 K, and the EES contribution gradually enhances for temperatures less than 100 K.

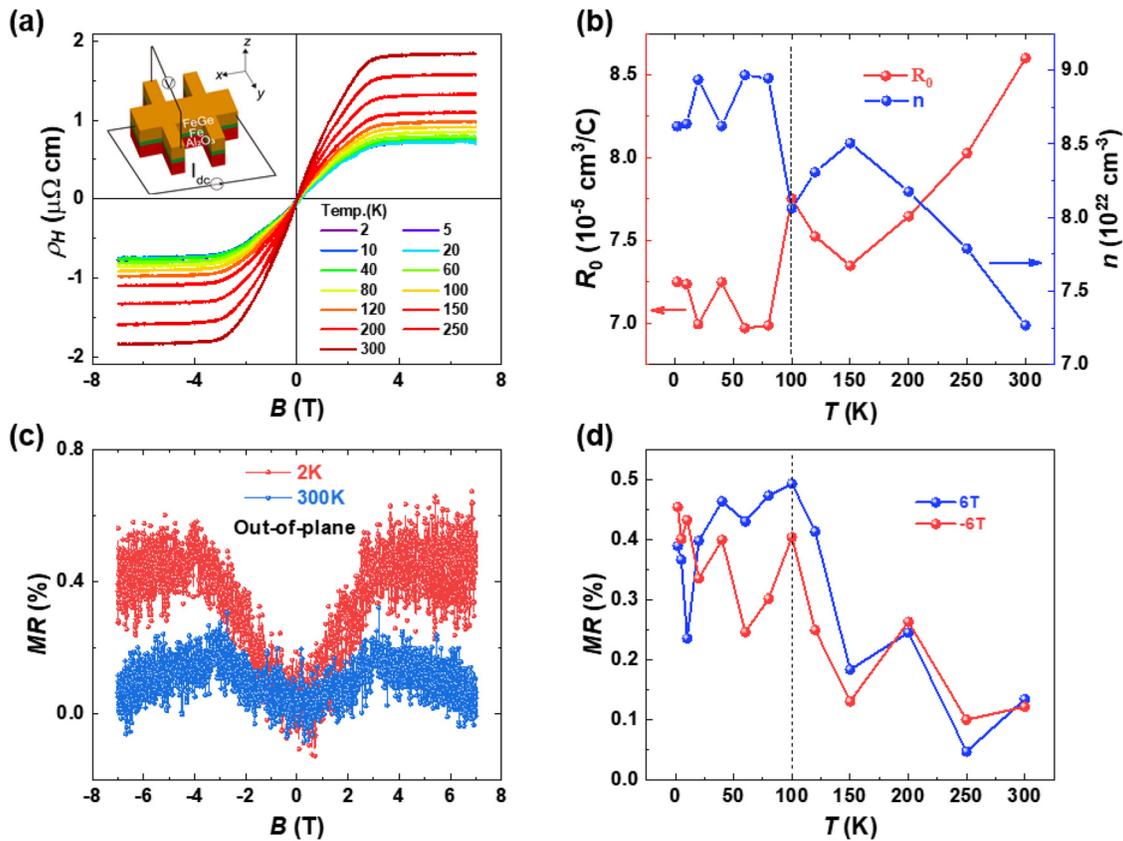

FIG. 3. (a) Hall resistivity as a function of out-of-plane magnetic field for Fe (2 nm)/FeGe (17 nm) thin film with temperature from 2 to 300 K. The inset presents a schematic diagram of the device testing setup. (b) Temperature dependence of ordinary Hall coefficient $R_0$ (red color) and corresponding carrier density $n$ (blue color). A sharp jump in $R_0$ and $n$ can be observed at around 100 K. (c) Magnetoresistance (MR) vs out-of-plane magnetic field measured at 2 K (red color) and 300 K (blue color). (d) Temperature dependence of MR under out-of-plane 6 T (blue color) and −6 T (red color). A rapid change in MR around 100 K can be also noticed, which would be related to the CDW in FeGe thin film.

To better understand the transport properties of our FeGe thin film, we conducted a piecewise fitting on the temperature dependence of resistivity. Interestingly, the $\rho_{xx}$ vs T curve can be divided into three parts and well fitted by different equations, respectively: $\rho = \rho_0 + a \times T$ for 100–380 K, $\rho = \rho_0 + a_1 \times T + a_2 \times T^2$ for 60–100 K, and $\rho = \rho_0 + b_1 \times T + b_2 \times T^2$ for 10–60 K, as shown in Figs. 2(b)–2(d). Three dominant scattering mechanisms including defect scattering, electron–phonon scattering (EPS), and electron–electron scattering (EES) were used to analyze our experimental data.[42–44] In the above equations, $\rho_0$ denotes the temperature-independent residual resistivity arising from defects and surface/interface scattering, $a_1 \times T$ and $a_2 \times T^2$ represent the EPS and EES contributions, respectively. Therefore, in the high temperature range of 380–100 K, the resistivity is governed almost entirely by EPS [inset of Fig. 2(b)], while the EES contribution gradually increases from 100 K, the fitting result involves both mechanisms, see Figs. 2(c) and 2(d), and EES plays a dominant role in the lower temperature range 60–10 K because $b_2$ is about three times larger than $a_2$ ($a_2 = 3.494 \times 10^{-4}$ $\mu\Omega$ cm/K$^2$, $b_2 = 9.458 \times 10^{-4}$ $\mu\Omega$ cm/K$^2$). Based on the experimental and fitted data, we observe a significant enhancement in the EES at temperatures below 100 K, which likely originates from the interaction between the CDW and magnetic order.

Subsequently, we focused on the magneto-transport characters of our Fe (2 nm)/FeGe (17 nm) thin film, here, the thin ferromagnetic Fe buffer layer would take important role. The Hall resistivity vs out-of-plane magnetic field, $\rho_H$ vs B, curve from 2 to 300 K is exhibited in Fig. 3(a). As the temperature rises, the saturation Hall resistivity gradually increases with $B > 3$ T. The anomalous Hall resistivity is mainly contributed by the Fe layer. From the equation $\rho_H = R_0 B + \rho_{AHE}$, here $R_0$ and $\rho_{AHE}$ are the ordinary Hall coefficient and anomalous Hall resistivity, respectively; $R_0$ was determined at different temperatures under $B > 3$ T. The $R_0$ vs T curve (red color) is plotted in Fig. 3(b). One notes a rapid jump at around 100 K (see the dashed line), correspondingly, the carrier density $n$ calculated from $R_0$ with temperature is summarized in Fig. 3(b) (see the blue colored curve). Importantly, these values of $R_0$ and $n$ were comparable to those of the kagome FeGe single crystal bulk.[45]

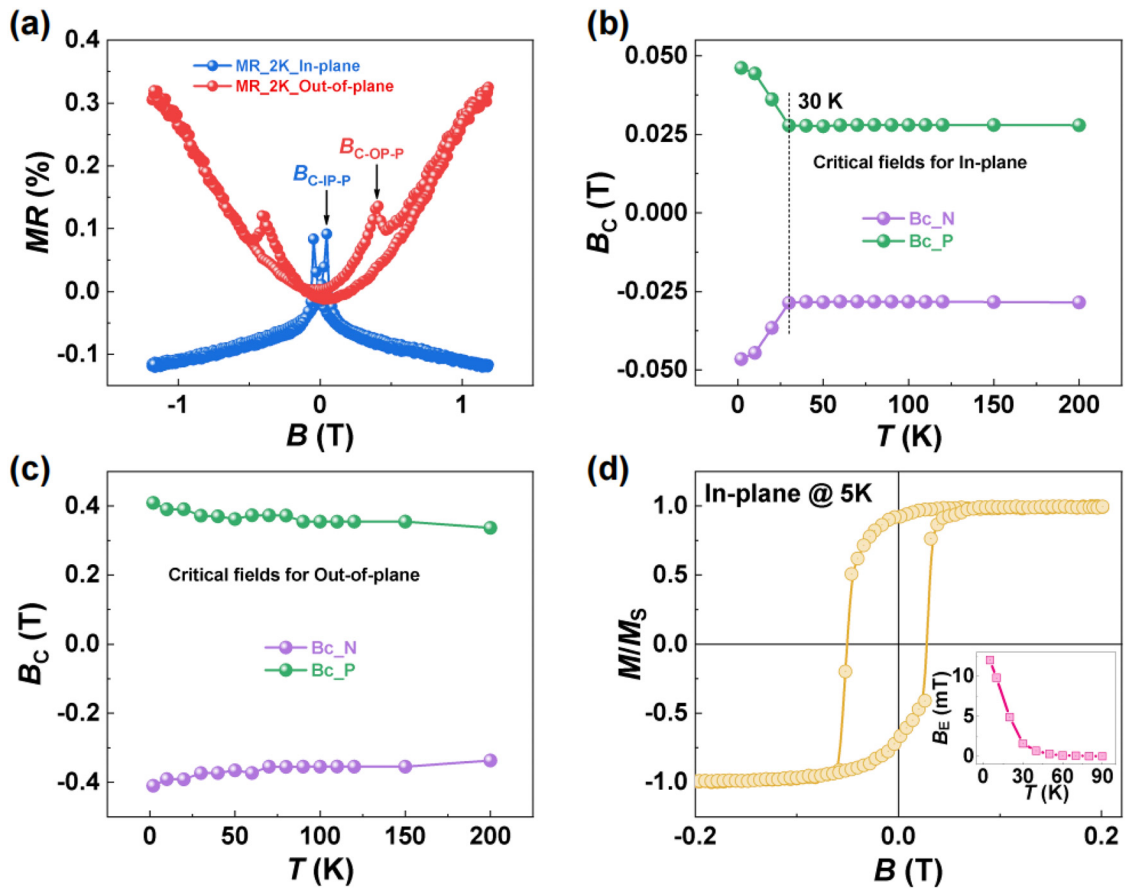

FIG. 4. (a) MR vs B curves under out-of-plane (red color) and in-plane (blue color) magnetic field measured at 2 K for Fe (2 nm)/FeGe (17 nm) thin film. The critical magnetic fields $B_{C-IP-P}$ and $B_{C-OP-P}$ for in-plane and out-of-plane conditions mainly depend on the contribution from 2-nm-thick Fe layer. (b) The in-plane critical magnetic field $B_C$ (negative $B_{C-N}$ and positive $B_{C-P}$) with temperature. The increase in absolute value of $B_C$ with $T < 30$ K (see the dashed line) could be due to the abrupt increase in canting angle of the magnetic moments. (c) Temperature dependence of $B_C$ (negative $B_{C-N}$ and positive $B_{C-P}$) for the out-of-plane condition. The absolute value of $B_C$ gradually increases with lowering temperature. (d) Normalized in-plane hysteresis loops of Fe (2 nm)/FeGe (17 nm) thin film measured at different temperatures after field cooling from 300 to 5 K at 0.1 T. Inset: the exchange bias field as a function of temperature.

We also investigated the magnetoresistance $\{MR = [\rho(B) - \rho(0)]/\rho(0) \times 100\%\}$ with temperature for our thin films. Figure 3(c) exhibits the MR vs B curves at 2 K (red color) and 300 K (blue color) under out-of-plane magnetic field. Positive MR at low magnetic field was observed at both temperatures, owing to the orbit-effect-induced behavior.[46] Moreover, the MR vs T curves under 6 and −6 T also present a sharp variation at around 100 K (see the dashed line), ascribed to the influence from the CDW in FeGe thin film.[45] Although the MR under high magnetic fields can be obtained, the low-field MR behavior is not clear [inset of Fig. 3(c)]. The interaction between Fe and FeGe layers under small magnetic field will better express the spin nature of the FeGe thin film. We thus measured the MR of our Fe (2 nm)/FeGe (17 nm) thin film using another low-field magneto-electrical measurement system.

As shown in Fig. 4(a), the MR vs B curves at 2 K display two sharp peaks for both in-plane and out-of-plane conditions, see the critical magnetic fields $B_{C-IP-P}$ and $B_{C-OP-P}$ indicated by the arrows. As the easy axis for Fe is in-plane, its in-plane coercivity is much smaller than the out-of-plane coercivity, resulting in a larger $B_C$ for the out-of-plane condition [Fig. 4(a)]. The temperature dependence of the in-plane $B_C$ (negative $B_{C-N}$ and positive $B_{C-P}$) is plotted in Fig. 4(b). On lowering the temperature, the absolute value of $B_C$ remains nearly constant but gradually increases from 30 K (see the dashed line), which coincides with the abrupt increase in the canting angle of the magnetic moments as reported in FeGe single crystal bulk.[18,24,25] For the out-of-plane condition, however, we did not observe such obvious change in MR; instead $B_C$ gradually increases with decreasing temperature as shown in Fig. 4(c). This phenomenon can be easily understood, as the Fe layer has in-plane easy axis, while the spins of kagome FeGe thin film are oriented out-of-plane [inset of Fig. 1(a)], resulting in weak Fe–FeGe interaction at high temperature. When the canting angle of the spins deviating from the c-axis increases at 30 K, the exchange coupling between Fe and FeGe is significantly enhanced, leading to a substantial increase in $B_C$. To verify this hypothesis, we measured the in-plane magnetic hysteresis loops (M–B curves) following field cooling of the sample from 300 to 5 K under an applied field of 0.1 T, as presented in Fig. 4(d). A distinct exchange bias effect is observed at low temperatures (particularly below 30 K), which is consistent with the results of our magneto-transport measurements. This result also demonstrates the high quality of our FeGe thin films.

In conclusion, we have demonstrated the epitaxial growth of kagome FeGe thin films using molecular beam epitaxy. The microstructure was characterized by XRD, AFM, and HR-STEM, and the transport properties were measured using PPMS and a low-field magneto-electrical measurement system. We found that a 2 nm-thick Fe buffer layer can not only be highly beneficial to the formation of hexagonal FeGe but also significantly enhance the flatness of the FeGe thin film. A Néel temperature of 397 K was obtained, which is slightly lower than that of the kagome FeGe single crystal bulk. Moreover, a sharp change in Hall coefficient and magnetoresistance around 100 K was also observed in the FeGe thin film, which might be ascribed to the effect of CDW. As the CDW is accompanied by a first-order structural transition characterized by strong dimerization along the c-axis of 1/4 of $Ge_1$-sites in $Fe_3Ge$ layers,[22,25] the thin film format offers an ideal platform to manipulate its structural variations via strain. In our work, it was found that despite the influence of the thin Fe buffer layer on the overall magneto-transport properties of Fe/FeGe thin films, scanning tunneling microscopy (STM) or angular resolved photoemission spectroscopy (ARPES) can be employed to investigate its surface properties and hence further understand the mechanism of CDW.

This work was supported by the National Key R&D Program of China (Grant No. 2022YFA1405100), the National Natural Science Foundation of China (Grant Nos. 12374125, 10225417, 12241405, 12427805, and 12474124), the NSAF (Grant No. U2330129), and the Chinese Academy of Sciences Project for Young Scientists in Basic Research (Grant No. YSBR-084).

## AUTHOR DECLARATIONS
### Conflict of Interest
The authors have no conflicts to disclose.

### Author Contributions
**Xiaoyue Song:** Writing – original draft (lead). **Yanshen Chen:** Data curation (equal). **Yongcheng Deng:** Formal analysis (equal). **Tongao Sun:** Data curation (equal). **Fei Wang:** Investigation (equal). **Guodong Wei:** Supervision (equal). **Xionghua Liu:** Writing – review & editing (equal). **Kaiyou Wang:** Project administration (equal).

## DATA AVAILABILITY
The data that support the findings of this study are available from the corresponding authors upon reasonable request.